\documentclass[12pt,a4paper]{article}
\usepackage{amsfonts,latexsym}
\usepackage{amsmath,amssymb}
\usepackage{graphicx,color}

\begin{document}

\vspace{10mm}

\begin{center}
{{{\Large {\bf Preheating  in the  nonminimal derivative coupling to gravity}}}}\\[10mm]

{Yun Soo Myung$^a$\footnote{e-mail address: ysmyung@inje.ac.kr} and Young-Jai Park$^b$\footnote{e-mail address:yjpark@sogang.ac.kr}}\\[8mm]

{$^a$Institute of Basic Sciences and Department  of Computer Simulation, Inje University Gimhae 50834, Korea\\[0pt]}
{$^b$Department of Physics, Sogang University, Seoul 04107, Korea\\[0pt]}

\vspace{5mm}

\end{center}
\vspace{2mm}

\begin{abstract}
We revisit the reheating mechanism after the end of inflation
in the non-minimal derivative coupling (NDC) to gravity with quadratic potential.
This is because the inflaton of the NDC should describe the slow-roll inflation
as well as  the preheating stage after the end of inflation.
We point out that the non-periodic inflaton solution implies the absence of parametric resonance,
compared to the periodic oscillating inflaton for the canonical coupling (CC) to gravity.
Furthermore, it is demonstrated that narrow and broad  parametric resonances do not appear
after the end of inflation in the NDC model by solving the differential equation numerically for the quantum field,
which differs from the case of the the CC model obtained by solving Mathieu equation.
\end{abstract}
\vspace{5mm}

{\footnotesize ~~~~PACS numbers: }


\newpage

\section{Introduction}

A period of accelerated expansion during the very early stage of the universe called inflation is able to account
for several otherwise difficult to explain features of the observed universe. A simplest inflation model is based on a
single slowly-rolling scalar field canonically coupled (CC) to gravity~\cite{muk}.
In the standard picture of the early universe, the universe passes through the period of reheating after the end of inflation.

The nonminimal derivative coupling (NDC)~\cite{Amendola:1993uh,Sushkov:2009hk} was firstly notified
by coupling the inflaton kinetic term to the Einstein tensor
such that the friction is gravitationally enhanced~\cite{Germani:2010gm}.
Later, this coupling has been  considered as an alternative  mechanism
to increase friction of an inflaton rolling down its own potential.
Actually, it makes a non-flat potential adequate for inflation without
introducing ghost
state~\cite{Germani:2011ua,Germani:2011mx}. This implies that during inflation,
the NDC increases friction, and flattens the potential effectively.

It is meaningful  to note that  there was a difference
between CC and NDC  even for taking the same quadratic  potential~\cite{Myung:2016twf}.
The difference appears clearly in the reheating process  after the end of inflation.
Reheating is being considered as an important part of inflationary universe
because it describes the production of Standard Model particles after the inflation~\cite{Allahverdi:2010xz}.
At this stage, the classical periodic  oscillating inflaton $\phi$ in the CC decays into massive bosons is
due to parametric resonance~\cite{Kofman:1994rk,Kofman:1996mv,Kofman:1997yn}.
In order to explain this phenomenon by introducing an interacting Lagrangian
of ${\cal L}_{\rm int}=-\frac{1}{2}g^2\phi^2\chi^2$,
the equation for quantum field  $\chi$ can be reduced to the Mathieu equation~\cite{Kofman:1994rk},
which is the well-known differential equation with periodic mass term
when neglecting the expansion of the universe.
This equation describes a harmonic oscillator with variable frequency (parametric oscillator).
In particular, if the coupling $g$ is large enough, the periodic modulation of the field mass
leads to strong instability via parametric resonance.

On the other hand, the inflaton in the NDC  oscillates with time-dependent frequency
which is surely a non-periodic function. The average solution $\phi$ in Eq. (\ref{ave-sol2})
mimics the  non-periodic nature of the inflation observed in Ref.~\cite{Myung:2016twf} numerically.
Therefore, the equation of quantum  field $\chi$  does not take a form of  the Mathieu equation,
and its solution  could not be obtained analytically.
However, the author in Ref. \cite{Ghalee:2013ada} has recently claimed
that the parametric resonance instability is absent,
by solving this equation with quadratic potential approximately.

In this work, we wish to revisit this important issue
because the inflaton of the NDC should describe the slow-roll inflation
as well as  the preheating stage after the end of inflation.

We will argue that there is no (narrow, broad) parametric resonance
after the end of inflation in the NDC model
because the field mass term is not a periodic function.
We will also numerically confirm it by solving the NDC-equation for $\chi_k$
and by comparing those obtained from the Mathieu equation.

\section{NDC with quadratic  potential}

Let us  consider  an inflation model  including the NDC of single scalar field
$\phi$ with the quadratic  potential~\cite{Myung:2016twf,Feng:2014tka,Myung:2015tga}
\begin{eqnarray} \label{mact}
S_{\rm}=\frac{1}{2}\int d^4x \sqrt{-g}\Big[M_{\rm
P}^2R+\frac{1}{\tilde{M}^2}G_{\mu\nu}\partial^{\mu}\phi\partial^{\nu}\phi-2V(\phi)\Big],~~V=\frac{m^2\phi^2}{2},
\end{eqnarray}
where $M_{\rm P}$ is a reduced Planck mass, $\tilde{M}$ is a mass
parameter and  $G_{\mu\nu}$ is the Einstein tensor. We find the CC model when replacing the second term by $-g_{\mu\nu}\partial^\mu\phi\partial^\mu\phi$.
Here we do not consider a conventional combination of CC+NDC[$-(g_{\mu\nu}-G_{\mu\nu}/\tilde{M}^2)\partial^\mu\phi\partial^\nu\phi$] in Ref.\cite{Tsujikawa:2012mk}
because this combination won't make the reheating analysis transparent.
From the action (\ref{mact}), one can easily derive  the Einstein and scalar(inflaton)
equations
\begin{eqnarray} \label{einseq}
&&G_{\mu\nu} =\frac{1}{M_{\rm P}^2} T_{\mu\nu},\\
\label{einseq-1}&&\frac{1}{\tilde{M}^2}G^{\mu\nu}\nabla_{\mu}\nabla_{\nu}\phi+m^2 \phi=0,
\end{eqnarray}
where $T_{\mu\nu}$ takes a complicated  form including fourth-order terms
\begin{eqnarray}
 T_{\mu\nu}&=&\frac{1}{\tilde{M}^2}\Big[\frac{1}{2}R\nabla_{\mu}\phi\nabla_{\nu}\phi
-2\nabla_{\rho}\phi\nabla_{(\mu}\phi R_{\nu)}^{\rho}
+\frac{1}{2}G_{\mu\nu}(\nabla\phi)^2-R_{\mu\rho\nu\sigma}\nabla^{\rho}\phi\nabla^{\sigma}\phi
\nonumber
\\&&\hspace*{5em}-\nabla_{\mu}\nabla^{\rho}\phi\nabla_{\nu}\nabla_{\rho}\phi
+(\nabla_{\mu}\nabla_{\nu}\phi)\nabla^2\phi\nonumber\\
&&
\label{em1}-g_{\mu\nu}\Big(\frac{\tilde{M}^2m^2}{2}\phi^2-R^{\rho\sigma}\nabla_{\rho}\phi\nabla_{\sigma}\phi+\frac{1}{2}(\nabla^2\phi)^2
-\frac{1}{2}(\nabla^{\rho}\nabla^{\sigma}\phi)\nabla_{\rho}\nabla_{\sigma}\phi
\Big)\Big].
\end{eqnarray}
Considering a  flat FRW spacetime  by introducing cosmic time $t$ as
\begin{eqnarray} \label{deds1}
ds^2_{\rm FRW}~=~\bar{g}_{\mu\nu}dx^\mu
dx^\nu~=~-dt^2+a^2(t)\delta_{ij}dx^idx^j, \label{deds2}
\end{eqnarray}
the first Friedmann
 and scalar equations  derived from Eqs. (\ref{einseq}) and (\ref{einseq-1}) are given
by
\begin{eqnarray}
H^2&=&\frac{1}{3M^2_{\rm P}}\rho_\phi=\frac{1}{3M_{\rm P}^2}\Big[\frac{9H^2}{2\tilde{M}^2}\dot{\phi}^2+\frac{1}{2}m^2\phi^2\Big],\label{Heq}\\
&&\hspace*{-4em}\frac{3H^2}{\tilde{M}^2}\ddot{\phi}+3H\Big(\frac{3H^2}{\tilde{M}^2}+\frac{2\dot{H}}{\tilde{M}^2}\Big)\dot{\phi}+m^2 \phi=0,\label{seq}
\end{eqnarray}
respectively~\cite{Myung:2016twf}.
Here $H=\dot{a}/a$ is the Hubble parameter and  the overdot
($\dot{}$) denotes  the differentiation with respect to time $t$. It is
evident from Eq. (\ref{Heq}) that the energy density $\rho_{\phi}$ for the NDC is
positive (ghost-free).

First of all, we would like to mention that  Qiu and Feng~\cite{Qiu:2016mrx} have recently suggested $a(t)$ and $H(t)$
during the rapid oscillation of the universe after inflation.
According to their approach, we wish to  look for a possible solution $\phi(t)$ with a quadratic potential
after the end of inflation by considering
\begin{equation} \label{sol}
a(t)\sim t^p,~~ H(t)=\frac{p}{t},~~\phi(t)=\Phi(t) \cos[q t^2],~~ \Phi(t)=\Phi_0t^{\frac{1-3p}{2}}.
\end{equation}
Taking into account Eq. (\ref{sol}), the scalar equation (\ref{seq}) can be solved exactly  for
\begin{equation}
q=\pm \frac{m\tilde{M}}{\sqrt{6}p},~~p=0,~\frac{1}{3},~\frac{5}{3}.
\end{equation}
Here we do not consider the case of $p=0$ because it is a trivial solution $\phi=0$ to Eq. (\ref{seq}).
Then, substituting Eq. (\ref{sol}) with the case of $q= \frac{m\tilde{M}}{\sqrt{6}p}$ and $p=\frac{5}{3}$ into
the right-hand side of Eq. (\ref{Heq}) leads to
\begin{equation}
\rho_\phi^{p=5/3}=\frac{m^2(7-5\cos[\frac{\sqrt{6}}{5} m\tilde{M} t^2])}{4t^4}+\frac{5\sqrt{6} m \sin[\frac{\sqrt{6}}{5} m\tilde{M}t^2]}{\tilde{M}t^6}+\frac{25(\cos[\frac{\sqrt{6}}{5} m\tilde{M} t^2]+1)}{\tilde{M}^2 t^8},
\end{equation}
which is not proportional to
\begin{equation} \label{hub-eq}
H^2=\frac{25}{9}\frac{1}{t^2}.
\end{equation}
Next, plugging Eq.  (\ref{sol}) with the case of  $q= \frac{m\tilde{M}}{\sqrt{6}p}$ and $p=\frac{1}{3}$ into   the right-hand side of Eq. (\ref{Heq}) leads to
\begin{equation}
\rho_\phi^{p=1/3}=\frac{m^2}{4}\Big(7-5\cos[\sqrt{6} m \tilde{M} t^2]\Big),
\end{equation}
which is also not obviously proportional to $H^2$ in Eq. (\ref{hub-eq}).
Note that although Eq. (\ref{sol}) is not  the   solution to Eq. (\ref{Heq}),
but it is surely  the   solution to Eq. (\ref{seq}).

On the other hand, following the averaging method~\cite{Ghalee:2013ada}
the other possible solution can be obtained by rewriting  Eqs. (\ref{Heq}) and (\ref{seq})
in terms of  $\tau$ defined as $t=\int(3H/m\tilde{M})d\tau$:
\begin{eqnarray}
&&H^2=\frac{m^2}{6M^2_{\rm P}}(\phi'^2+\phi^2), \label{neweq1}\\
&&\phi''+\Big(\frac{H'}{H}+\frac{9H^2}{m\tilde{M}}\Big)\phi'+3\phi=0, \label{neweq2}
\end{eqnarray}
where the prime($'$) denotes the differentiation with respect to $\tau$.
Then, the parametric solution to Eq. (\ref{neweq1}) could be assumed to be
\begin{eqnarray} \label{nphi1}
\phi&=&\frac{\sqrt{6}HM_{\rm P}}{m} \sin[\tau+f(\tau)],\\
\phi'&=&\frac{\sqrt{6}HM_{\rm P}}{m} \cos[\tau+f(\tau)],\label{nphi2}
\end{eqnarray}
where $f(\tau)$ is an arbitrary function.
Noting that the differentiation of Eq. (\ref{nphi1}) should be Eq. (\ref{nphi2}), one finds that
\begin{equation} \label{h-eq}
Hf'\cos[\tau+f(\tau)]+H'\sin[\tau+f(\tau)]=0.
\end{equation}
Then, the differenciation of $\phi'$ is given by
\begin{equation} \label{nphi3}
\phi'' = - (1+f')\phi +(H'/H) \phi' .
\end{equation}
Plugging Eqs. (\ref{nphi1}), (\ref{nphi2})  and (\ref{nphi2}) into Eq. (\ref{neweq2}) leads to
\begin{equation} \label{f-eq}
\Big(2H'+\frac{9H^3}{m\tilde{M}}\Big)\cos[\tau+f(\tau)]+(2H-Hf')\sin[\tau+f(\tau)]=0.
\end{equation}
Then, solving Eqs. (\ref{h-eq}) and (\ref{f-eq}) leads to
\begin{eqnarray}
f'&=&\frac{\sin[\tau+f(\tau)]}{1+\cos^2[\tau+f(\tau)]}\Big\{2\sin[\tau+f(\tau)]+\frac{9H^2}{m\tilde{M}}\cos[\tau+f(\tau)]\Big\},\label{fp-eq}\\
H'&=&-\frac{H\cos[\tau+f(\tau)]}{1+\cos^2[\tau+f(\tau)]}\Big\{2\sin[\tau+f(\tau)]+\frac{9H^2}{m\tilde{M}}\cos[\tau+f(\tau)]\Big\}.\label{hp-eq}
\end{eqnarray}
However, it is almost impossible to solve Eqs. (\ref{fp-eq}) and (\ref{hp-eq}) because $f(\tau)$ is an argument of  $\cos$- and $\sin$-functions.
At this stage, we note  that
an approximate  solution might  be obtained when introducing the averaging
method~\cite{Ghalee:2013ada}. In this case,
one finds the averaged differential equations
\begin{equation}
f'_{\rm a}=2(\sqrt{2}-1),~~ H'_{\rm a}=-\frac{9H^3_{\rm a}}{m\tilde{M}}\Big(1-\frac{1}{\sqrt{2}}\Big).
\end{equation}
Here the subscript (a) implies an average over $\tau (f'_{\rm a}=\int^\pi_0 f'(\tau)d\tau/\pi$)
for the fast varing quantities $f'$ and $H$ by assuming that the slowy varing quantities $H$ and $f$ are fixed.
As a result, we easily obtain the desired solutions
\begin{equation}
f_{\rm a}(\tau)=2(\sqrt{2}-1)\tau,~~ H_{\rm a}(\tau)=\frac{1}{\sqrt{\frac{18}{m\tilde{M}}(1-\frac{1}{\sqrt{2}})\tau}}.
\end{equation}
Moreover, we may derive the relation between $t$ and $\tau$ as
\begin{equation}
t=\frac{2 \sqrt{\tau}}{\sqrt{2m\tilde{M}(1-\frac{1}{\sqrt{2}})}}
\end{equation}
which explicitly shows the behavior of $\tau \propto t^2$.
This is origin of why $\phi(t)$ has the non-periodic solution after the end of inflation~\cite{Myung:2016twf}.

Finally, we obtain the average solutions for $H$ and $\phi$ as
\begin{eqnarray}
&&a_{\rm a}\propto t^{\frac{2}{3(2-\sqrt{2})}},~H_{\rm a}(t)=\frac{2}{3(2-\sqrt{2})t},\label{ave-sol1}\\
&&\phi_{\rm
a}(t)\equiv \Phi_{\rm a}(t)\sin\Big[\tilde{m}^2t^2\Big]= \frac{\sqrt{6}M_{\rm P}H_{\rm a}(t)}{ m}
\sin\Big[\tilde{m}^2t^2\Big], \label{ave-sol2} \\
&&\dot{\phi}_{\rm
a}(t)= \frac{\sqrt{6}M_{\rm P}\tilde{M}}{3}
\cos\Big[\tilde{m}^2t^2\Big]  \label{ave-sol3}
\end{eqnarray}
with
\begin{equation}
\tilde{m}^2=\frac{m\tilde{M}}{2}(2-\sqrt{2})(\sqrt{2}-\frac{1}{2}).
\end{equation}
Importantly, we observe that  $\phi_{\rm a}(t)$ oscillates with time-dependent frequency.
As a result, it is not a periodic function of $t$.
It is worth to note that Eqs. (\ref{ave-sol2}) and (\ref{ave-sol3}) satisfy the first Friedmann equation (\ref{Heq}) exactly,
while  they satisfy the scalar equation (\ref{seq}) approximately  for large $t$.
Hence, we may  regard  Eqs. (\ref{ave-sol1})-(\ref{ave-sol3}) as the best analytic solution which describes the reheating process
after inflation in the NDC.
We note that the approximate solution found in~Ref. \cite{Ghalee:2013ada} takes the  forms (\ref{ave-sol1})-(\ref{ave-sol3}) obtained
when replacing $\sin[\tilde{m}^2t^2]$ in Eq. (\ref{ave-sol2}) and $\cos[\tilde{m}^2t^2]$ in Eq. (\ref{ave-sol3})
by $\cos[\tilde{m}^2t^2]$ and $-\sin[\tilde{m}^2t^2]$, respectively.

In conclusion, the non-periodic nature of the inflation is regarded as a clear feature of reheating process in the NDC,
when one compares it with the periodic inflation of $\phi(t)\propto\sin(mt)/t$ in the CC~\cite{Kofman:1994rk,Kofman:1996mv,Kofman:1997yn}.
The solution (\ref{ave-sol2}) mimics the  non-periodic nature of the inflation observed in~\cite{Myung:2016twf} numerically.
Furthermore, Eq. (\ref{ave-sol3}) indicates that the velocity of inflaton $\dot{\phi}$  oscillates without
damping for the NDC~\cite{Myung:2015tga}, while it oscillates with damping for the CC[$\dot{\phi}(t)\propto\cos(mt)/t$ in~\cite{Ghalee:2013ada}].
We are ready for studying the parametric resonance because we are aware of an analytic form for the inflaton (\ref{ave-sol2}) in the reheating.

\section{Parametric resonance}

It was reported   that the parametric resonance is absent for
NDC when considering the decay of the scalar field $\phi$ into a quantum  field $\chi$~\cite{Ghalee:2013ada}, whereas the
parametric resonance is present for CC. However, this statement is not clear.
In this section, we wish to revisit this issue.

Now, let us consider the relevant quantum field Lagrangian is given by
\begin{equation} \label{chi-lag}
{\cal L}_{\chi}=-\frac{1}{2}\partial_\mu\chi \partial^\mu\chi-\frac{1}{2} m^2_\chi\chi^2-\frac{1}{2}g^2\phi^2\chi^2.
\end{equation}
The time evolution of the quantum fluctuation $\chi$ is governed by the classical equation of motion in the flat FRW universe (\ref{deds1}) as
\begin{equation} \label{KG-eq}
\ddot{\chi}+3H\dot{\chi}-\frac{1}{a^2} \nabla^2_x \chi +g^2\phi^2 \chi=0.
\end{equation}
With $\vec{x}$ and $\vec{k}$ representing the comoving position and momentum vectors, $\chi$ can be expressed as
\begin{equation} \label{chi-exp}
\chi(t,\vec{x})=\int \frac{d^3k}{(2\pi)^{3/2}} \Big[a_k\chi_k(t)e^{-i \vec{k}\cdot \vec{x}}+a_k^\dagger\chi^*_k(t)e^{i \vec{k}\cdot \vec{x}}\Big],
\end{equation}
where $a_k$ and $a_k^\dagger$ are annihilation and creation operators, respectively. We assume $m^2_\chi=0$ for simplicity.
Plugging (\ref{chi-exp}) into (\ref{KG-eq}) leads to the equation for temporal part of the Fourier mode  $\chi_k$  with $k=|\vec{k}|$
\begin{equation} \label{KGF1-eq}
\ddot{\chi}_k+3H\dot{\chi}_k+\Big(\frac{k^2}{a^2} +g^2\phi^2(t)\Big) \chi_k=0.
\end{equation}
In order to promote a further computation, let us ignore the expansion of the universe and
assume a slow variation of $\Phi_{\rm a}(t)$ compared to oscillation frequencies of the fields $\phi$ and $\chi$:
\begin{equation}
H_{\rm a}\approx 0;~a_{\rm a}\approx1;~\Phi_{\rm a}(t)\approx{\rm const}.
\end{equation}
If the coupling $g$ is large enough, one can ignore the friction term $3\dot{H} \chi_k$.
Then, Eq. (\ref{KGF1-eq}) takes the form
\begin{equation} \label{KGF2-eq}
\ddot{\chi}_k+\Big(k^2 +g^2\Phi_{\rm a}^2\sin^2(\tilde{m}^2t^2)\Big) \chi_k=0.
\end{equation}
Let us define  new variables like as
\begin{equation}
\xi=\sqrt{2}\tilde{m} t,~~\omega_k^2=\frac{k^2}{2 \tilde{m}^2}+2\tilde{\epsilon},~~ \tilde{\epsilon}=\frac{g^2\Phi^2_{\rm a}}{8\tilde{m}^2}.
\end{equation}
Then,  Eq. (\ref{KGF1-eq}) leads to the differential equation with the non-periodic mass term as
 \begin{equation} \label{KGF3-eq}
\frac{d^2\chi_k}{d\xi^2}+\Big[\omega^2_k-2\tilde{\epsilon} \cos(\xi^2)\Big] \chi_k=0,
\end{equation}
which is called the NDC-differential equation.

On the other hand, the CC case with quadratic potential arrives at the Mathieu equation
with $z=mt$~\cite{Kofman:1994rk,Kofman:1997yn}
\begin{equation} \label{KGF4-eq}
\frac{d^2\tilde{\chi}_k}{dz^2}+\Big[A_k-2\tilde{q} \cos(2z)\Big] \tilde{\chi}_k=0
\end{equation}
with
\begin{equation} \label{anq}
 A_k=\frac{k^2}{m^2}+2\tilde{q},~~ \tilde{q}=\frac{g^2\Phi^2}{4m^2}.
 \end{equation}
Here $\tilde{\chi}_k$ of the CC case is a quantum field corresponding to ${\chi}_k$ of the NDC case.
If the coupling $g$ is large enough, periodic modulation of the field mass leads to strong instability via parametric resonance.
According to Floquet's theorem, a general solution to the Mathieu equation (\ref{KGF4-eq}) takes the form
\begin{equation} \label{M-sol}
\tilde{\chi}_k(z)=e^{\mu z} P(z),
\end{equation}
where $P(z)$ is a periodic function with period $\pi$. The Floquet exponent $\mu(A_k,\tilde{q})$ depends on parameters $A_k$ and $\tilde{q}$.
In the case of positive Re[$\mu$], one has an exponential instability of the solution.
The growth of the mode $\tilde{\chi}_k$ corresponds to particle production,
as in the case of particle production in the external gravitational field.
The case of $\tilde{q} <1$ leads to the narrow parametric resonance,
while the case of $\tilde{q}>1$ provides the broad parametric  resonance.

At this stage, we should admit  that one could not solve the non-periodic differential equation (\ref{KGF3-eq}) with $\epsilon=2\tilde{\epsilon}$ directly.
To find   an approximate solution, one  may expand $\chi_k(\xi)$ as
\begin{equation} \label{chis}
\chi_k=\chi^0_k+\epsilon \chi^1_k +\cdots.
\end{equation}
The author in Ref. \cite{Ghalee:2013ada} has  argued that the expression (\ref{chis}) is valid
if $\chi^1_k$ has no terms which grow without bound as $\xi \to \infty$.
Plugging (\ref{chis}) into (\ref{KGF4-eq}) and keeping all terms up to second-order in $\epsilon$,
one finds two equations
\begin{eqnarray} \label{zeroth-eq}
&& \frac{d^2\chi^0_k}{dz^2}+\omega^2_k \chi_k^0=0, \\
&&  \frac{d^2\chi^1_k}{dz^2}+\omega^2_k \chi_k^1=\chi^0_k\cos(\xi^2).
 \label{first-eq}
\end{eqnarray}
The homogeneous equation (\ref{zeroth-eq}) of zeroth order yields a general solution
\begin{equation}
\chi^0_k=b_1 \sin(\omega_k \xi)+b_2 \cos(\omega_k \xi).
\end{equation}
Introducing new variables $u/v\equiv\sqrt{2/\pi}(\xi\pm\omega_k)$, the inhomogeneous equation (\ref{first-eq}) of first order can be solved to give
\begin{eqnarray}
\chi^1_k&=&c_1\sin(\omega_k \xi)+c_2 \cos(\omega_k \xi) \nonumber \\
&+&\frac{1}{4\omega_k}\sqrt{\frac{\pi}{2}}\Big\{ \cos(\omega_k\xi+\omega^2_k)[b_1C(u)-b_2S(u)]+ \sin(\omega_k\xi+\omega^2_k)[b_1S(u)+b_2C(u)] \nonumber \\
&+&  \cos(\omega_k\xi-\omega^2_k)[b_1C(v)+b_2S(v)]+\sin(\omega_k\xi-\omega^2_k)[-b_1S(v)+b_2C(v)]\nonumber \\
&+& 2C((u+v)/2)[-b_1\cos(\omega_k\xi)+b_2\sin(\omega_k\xi)]\Big\}, \label{inh-sol}
\end{eqnarray}
where Fresnel-cosine integral $C(u)$ and Fresnel-sine integral $S(u)$ are defined  by
\begin{equation}
C(u)=\int^u_0\cos\Big[\frac{\pi}{2} x^2\Big]dx,~~S(u)=\int^u_0\sin\Big[\frac{\pi}{2} x^2\Big]dx,
\end{equation}
respectively.
These are surely non-periodic finite functions because $\lim_{u\to \infty}C(u)=1/2$ and $\lim_{u\to \infty}S(u)=1/2$.
One may attempt to conclude that the parametric resonance is absent in Eq. (\ref{KGF3-eq}) because $\chi^1_k$ (\ref{inh-sol}) does not have any terms which grow without bound as $\xi(u,v)\to \infty$.
However, in deriving this approximate solution, Ghalee~\cite{Ghalee:2013ada} has neglected $\epsilon$ itself  in Eq. (\ref{KGF3-eq}) which plays an important role in determining its solution. Hence, we insist that the solution (\ref{inh-sol}) is not the correct one.

In order to support it, one may  rewrite  the Mathieu equation (\ref{KGF4-eq}) in terms of $t$ as
 \begin{equation}
 \frac{d^2 \tilde{\chi}_k}{dt^2}+[\omega^2-\varepsilon \cos(2mt)]\tilde{\chi}_k=0,~~\omega^2=k^2+\varepsilon,~~\varepsilon=\frac{g^2\Phi^2}{2}.
 \end{equation}
 Introducing
 \begin{equation} \label{chisCC}
\tilde{\chi}_k=\tilde{\chi}^0_k+\varepsilon \tilde{\chi}^1_k +\cdots,
\end{equation}
its equations are given by
\begin{eqnarray} \label{CCzeroth-eq}
&& \frac{d^2\tilde{\chi}^0_k}{dt^2}+\omega^2\tilde{ \chi}_k^0=0, \\
&&  \frac{d^2\tilde{\chi}^1_k}{dt^2}+\omega^2 \tilde{\chi}_k^1=\tilde{\chi}^0_k\cos(2mt).
 \label{CCfirst-eq}
\end{eqnarray}
An  approximate  solution is given by
 \begin{eqnarray}
 \tilde{\chi}_k^0(t)&=&  b\sin(\omega t)+c \cos(\omega t), \label{cc-sol1} \\
 \tilde{\chi}_k^1(t)&=&\tilde{c}_1 \cos(\omega t)+\tilde{c}_2 \sin(\omega t) \nonumber \\
 &+&\frac{1}{4m(m^2-\omega^2)}\Big\{\cos(\omega t)[b\omega \sin(2mt)+c m\sin^2(mt)]\nonumber \\
 &+&\sin(\omega t)[bm \sin^2(mt)-c \omega\sin(2mt)]-m \cos^2(mt)[b\sin(\omega t)+c \cos(\omega t)]\Big\}. \label{cc-sol2}
 \end{eqnarray}
We stress to note that $\tilde{\chi}^1_k(t)$ is an oscillating function for any $t$,
and it does not blow up unless $m=\omega$.
This contradicts to the solution (\ref{M-sol}) to the Mathieu equation.
Hence this approach to obtaining  approximate  solutions (\ref{cc-sol1}) and (\ref{cc-sol2}) could  not be trusted.

\section{Numerical analysis: no parametric resonance}
\begin{figure*}[t!]
\centering
\includegraphics[width=.5\linewidth,origin=tl]{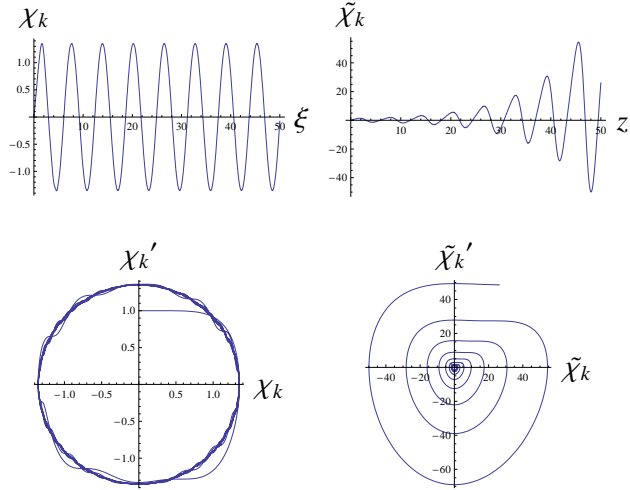}
\caption{Top-left:  oscillating mode $\chi_k$ for $\omega^2_k=1$ and $ \tilde{\epsilon}=0.5$ in the NDC, and top-right: growth of $\tilde{\chi}_k$ for $A_k=1$ and $\tilde{q}=0.5$ in the CC.
Bottom-left: parametric plot for ($\chi_k,\chi_k'$) in the NDC where the prime denotes the derivative with respect to $\xi$, and bottom-right: parametric plot for ($\tilde{\chi}_k,\tilde{\chi}_k'$) in the CC where the prime denotes the derivative with respect to $z$.}
\end{figure*}
\begin{figure*}[t!]
\centering
\includegraphics[width=.5\linewidth,origin=tl]{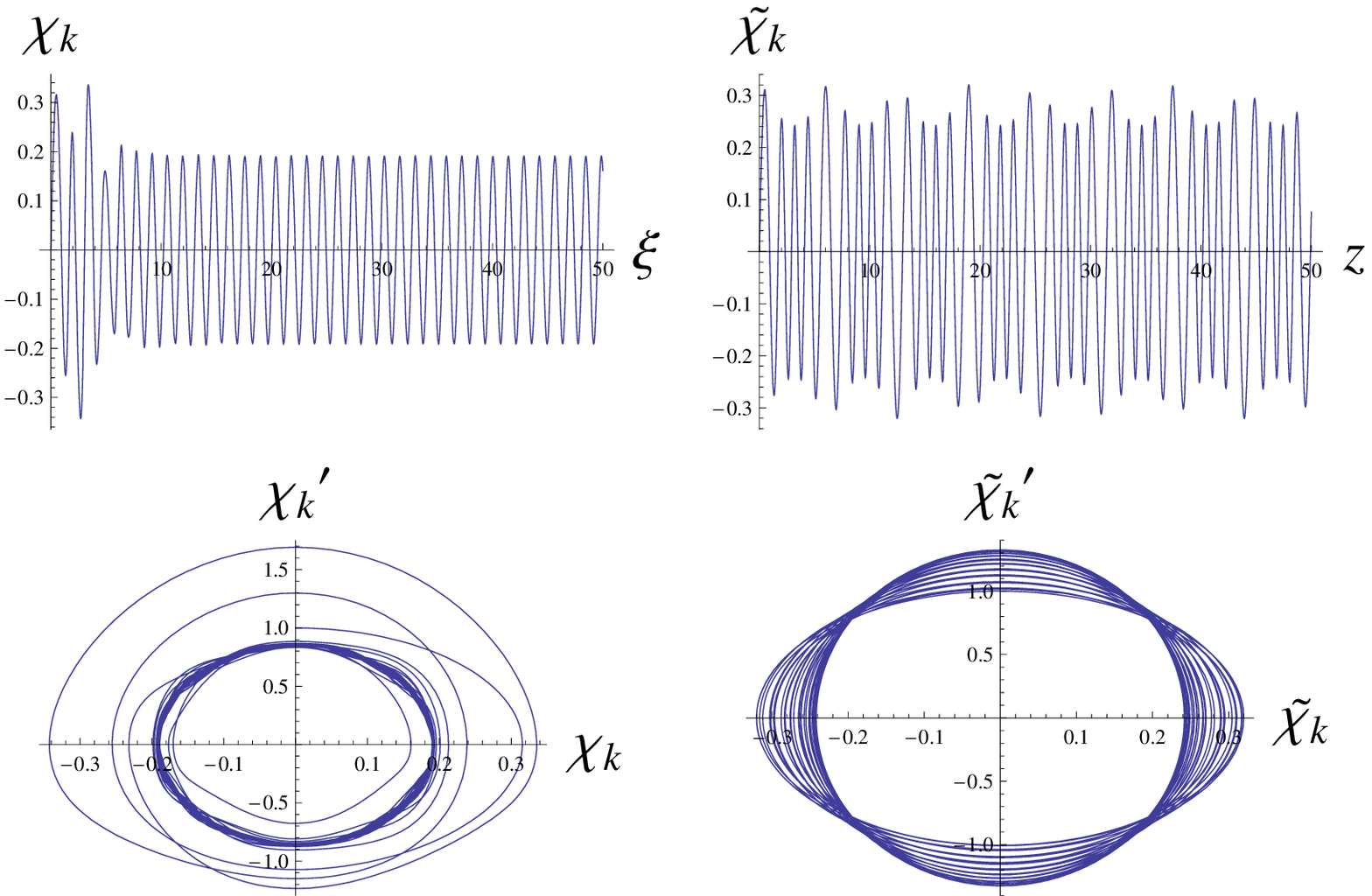}
\caption{Top-left:  oscillating mode $\chi_k$ for $\omega^2_k=20$ and $\tilde{\epsilon}=5$ in the NDC, and top-right:  oscillating mode of  $\tilde{\chi}_k$ for $A_k=20$ and $\tilde{q}=5$ in the CC.
Bottom-left: parametric plot for ($\chi_k,\chi_k'$) in the NDC, and bottom-right: parametric plot for ($\tilde{\chi}_k,\tilde{\chi}_k'$) in the CC. }
\end{figure*}
\begin{figure*}[t!]
\centering
\includegraphics[width=.5\linewidth,origin=tl]{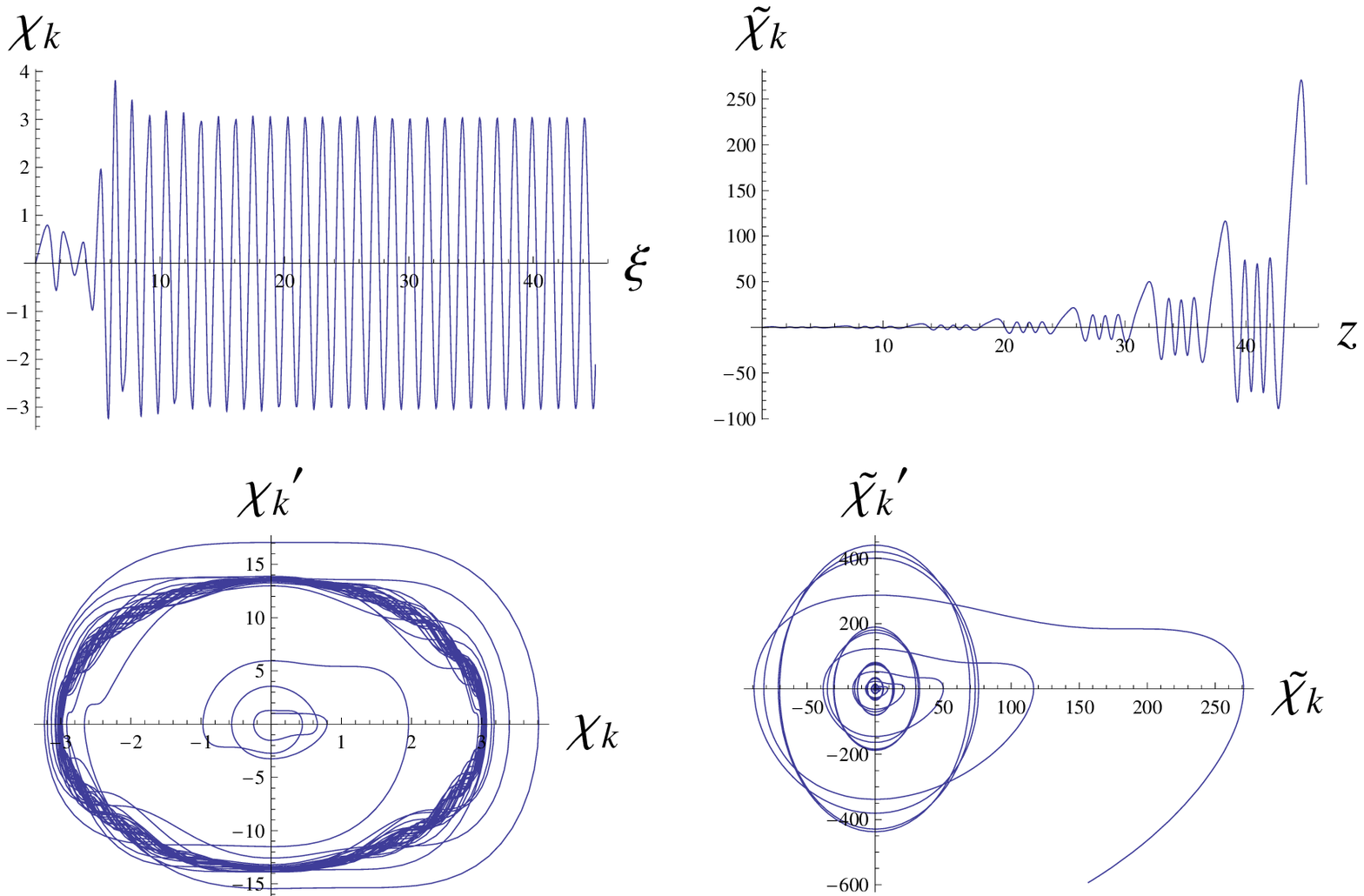}
\caption{Top-left: oscillating  mode $\chi_k$ for $\omega^2_k=20$ and $\tilde{\epsilon=}10$ in the NDC, and top-right: growth of $\tilde{\chi}_k$ for $A_k=20$ and $\tilde{q}=10$ in the CC.
Bottom-left: parametric plot for ($\chi_k,\chi_k'$) in the NDC,  and bottom-right: parametric plot for ($\tilde{\chi}_k,\tilde{\chi}_k'$) in the CC. }
\end{figure*}
\begin{figure*}[t!]
\centering
\includegraphics[width=.5\linewidth,origin=tl]{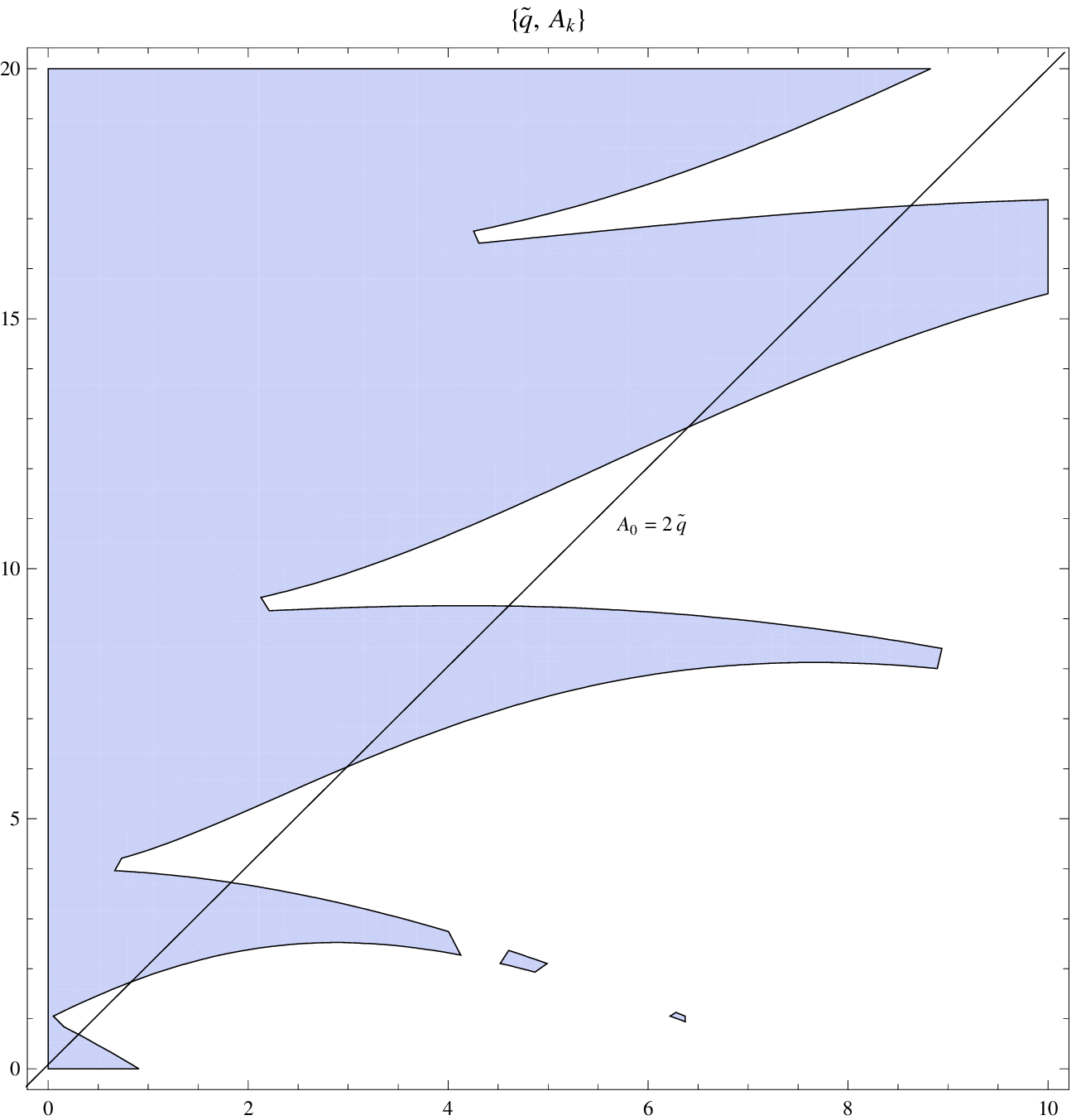}
\caption{Sketch of the stability-instability chart of  the Mathieu equation. Gray bands indicate regions of stability, while white bands denote region of instability.
The line $A_0=2\tilde{q}$ shows the values of $A_k$ and $\tilde{q}$ for $k=0$.  We choose $A_k=1,20$ for comparison test.  }
\end{figure*}
In this section, let us numerically solve the NDC-equation (\ref{KGF3-eq}),
and also solve Mathieu equation (\ref{KGF4-eq}) of the CC case in order to compare to the NDC one.
We observe  inequalities of $\omega^2_k\ge 2\tilde{\epsilon}$ and $A_k \ge 2\tilde{q}$.
For this purpose, we choose three proper cases: \\
(i) $\omega^2_k=A_k=1$ and $\tilde{\epsilon}=0.5,~\tilde{q}=0.5$, \\
(ii) $\omega^2_k=A_k=20$ and $\tilde{\epsilon}=5,~\tilde{q}=5$, \\
(iii) $\omega^2_k=A_k=20$ and $\tilde{\epsilon}=10,~\tilde{q}=10$. \\
As was shown in Fig. 1[(i) case], we find  that the top-left corresponds to the homogeneous oscillating mode for the NDC,
whereas the top-right is a growth mode which  represents the narrow parametric resonance for the CC ($\tilde{q}<1$).
The bottom-left and -right confirm the oscillating and  growing  modes, respectively.
The case (ii)[Fig. 2] does not show the difference between NDC and CC significantly because two belong to oscillating modes.
We observe from Fig. 3 that the case (iii) indicates clearly that the top-left denotes a oscillating mode for the NDC,
 while the top-right is a rapidly growing mode which  represents broad parametric resonance for the CC ($\tilde{q}>1$).
Moreover, in the case of CC, the resonance is much more efficient if $\tilde{q} \gg 1$.
These observations for the CC  could be confirmed from Fig. 4, which shows the stability-instability chart of the Mathier equation.
In addition, we have solved the NDC-equation (\ref{KGF3-eq}) for different $\omega^2_k$ and $\tilde{\epsilon}$ numerically,
and compared those obtained from different sets of $A_k$ and  $ \tilde{q}$.
We have also arrived at the same result.

\section{Discussions}
If the NDC is a promising coupling for obtaining a successful inflation rather than the CC,
the inflaton of the NDC should describe the slow-roll inflation as well as  the preheating stage after the end of inflation.
It turned out that  this coupling has been  considered as an alternative  mechanism  to increase friction of an inflaton rolling down its
own potential~\cite{Germani:2011ua,Germani:2011mx}.

However, after the end of inflation, the inflaton in the NDC  oscillates with
time-dependent frequency which is surely  a non-periodic function.
The solution of the form (\ref{ave-sol2}) mimics the non-periodic nature of the inflation
observed in~\cite{Myung:2016twf} numerically,
while  Eq. (\ref{ave-sol3}) dictates that the velocity of inflaton $\dot{\phi}$ oscillates
without damping for the NDC~\cite{Myung:2015tga}.

In order to see whether the parametric resonance occurs or not in the NDC,
we have introduced the Lagrangian (\ref{chi-lag}) for the quantum field $\chi$.
The differential equation (\ref{KGF3-eq}) of quantum mode $\chi_k$ did not take a form of the Mathieu equation (\ref{KGF4-eq})
with periodic mass term, and thus its solution could not be obtained analytically.
First, we have argued  that there is no (narrow, broad) parametric resonance after the end of inflation in the NDC model
because the field mass term is not a periodic function.
We have also numerically confirmed it by solving the NDC-equation (\ref{KGF3-eq}) for $\chi_k$ and
by comparing those obtained from the Mathieu equation (\ref{KGF3-eq}).

 \vspace{0.25cm}
  {\bf Acknowledgement}

\vspace{0.25cm}
 This work was  supported by  the 2016 Inje University research grant.

\end{document}